# Interfacial Load Monitoring and Failure Detection in Total Joint Replacements via Piezoresistive Bone Cement and Electrical Impedance Tomography

H. Ghaednia, C. E. Owens, R. Roberts, T. N. Tallman, A. John Hart, and K. M. Varadarajan

*Abstract*— Aseptic loosening, or loss of implant fixation, is a common complication following total joint replacement. Revision surgeries cost the healthcare system over $8 billion annually in the US. Despite the prevalence of aseptic loosening, timely and accurate detection remains a challenge because traditional imaging modalities such as plain radiographs struggle to reliably detect the early stages of implant loosening. Motivated by this challenge, we present a novel approach for in vivo monitoring and failure detection of cemented joint replacements. Poly(methyl methacrylate) (PMMA) bone cement is modified with low volume fractions of chopped carbon fiber (CF) to impart piezoresistive-based self-sensing. Electrical impedance tomography (EIT) is then used to detect and monitor load-induced deformation and fracture of CF/PMMA in a phantom tank. We therefore show that EIT indeed is able to adeptly detect loading force on a prosthetic surrogate, distinguish between increasing load magnitudes, detect failure of implant fixation, and even distinguish between cement cracking and cement debonding. Because EIT is a low-cost, physiologically benign, and potentially real-time imaging modality, the feasibility study herein presented has potential to transform the standard of care for post-operative monitoring of implant conditions. Beyond clinical relevance, this technique could positively impact orthopedic researchers by providing, via in vivo monitoring, insight into the factors that initiate aseptic loosening.

*Index Terms*— electrical impedance tomography, piezoresistive, total joint replacement, aseptic loosening,

Financial support was provided by the MGH/MIT Strategic Partnership Grand Challenge Grant (to K.M.V. and A.J.H.), the MIT - Tecnológico de Monterry Nanotechnology Program (to R.R. and A.J.H.), and the Department of Defense (DoD) though the National Defense Science and Engineering Graduate (NDSEG) program (to C.E.O.).

H. G. is with Department of Orthopaedic Surgery, Massachusetts General Hospital, Boston, MA 02114 (e-mail: hghaednia@mgh.harvard.edu).
C. E. O. is with Department of Mechanical Engineering, Massachusetts Institute of Technology, Cambridge, MA 02139 (e-mail: crystalo@mit.edu).
R. R. is with School of Engineering and Sciences, Tecnologico de Monterrey, Mexico 64849 (e-mail: robertsu@mit.edu).
T. N. T. is School of Aeronautics and Astronautics, Purdue University, West Lafayette, IN 47907 (e-mail: ttallman@purdue.edu).
A. J. H. is with Department of Mechanical Engineering, Massachusetts Institute of Technology, Cambridge, MA 02139 (e-mail: ajhart@mit.edu).
K. M. V. is with Department of Orthopaedic Surgery, Massachusetts General Hospital, Boston, MA 02114 (e-mail: kmangudivaradarajan@mgh.harvard.edu).

## I. INTRODUCTION

KNEE, hip, shoulder, and other joint replacement procedures benefit over two million patients each year worldwide [1] [2] [3] [4]. However, failures of these surgeries pose significant risks to patient health and present an enormous economic burden to the healthcare system; for instance, revisions in the United States cost over $8 billion per year [5] [6]. Loss of implant fixation, called aseptic loosening, is the leading cause of revision following primary joint replacement and accounts for approximately 25% of all revision cases [7] [8]. In particular, patients younger than 55 years of age face an elevated risk of revision due to the greater demands placed on their joints and due to the steady increase in risk of implant loosening with in vivo duration.

Despite the clinical need for detection of implant loosening, definitive diagnosis is challenging. Plain radiographs, the gold standard for clinical diagnosis, are largely limited to detection of aseptic loosening in advanced stages, with relevance of early radiolucent lines remaining a subject of debate [9] [10]. Further, patient follow up over an extended period is often needed before definitive diagnoses of implant loosening can be made. The importance of this clinical issue is perhaps best illustrated by example. In early 2017, Bonutti et al. reported high rates of early tibial component loosening in patients with a new implant design (15 knees, < 2 years post-surgery) [11]. Patients presented to the physician's office with complaints of pain upon weight-bearing and underwent physical examination, radiographic evaluation, and screening for joint infection. Radiographic evaluation demonstrated loosening in only 2 out of the 15 knees. However, intra-operatively all knees were found to have grossly loose implants requiring revision. Other techniques such as computed tomography and arthrography can increase diagnostic accuracy but are more expensive, expose patients to higher doses of ionizing radiation, and involve risks related to use of contrast agents. Furthermore, pain, which is the primary symptom of a loose implant, is also a symptom of peri-prosthetic joint infection. The above discussion demonstrates the critical importance of early, cost-effective, and accurate diagnosis of implant loosening to prevent serious harm to large groups of patients. A diagnostic tool that can meet this need would also allow us to better understand and ultimately address the surgical, patient, and implant factors responsible for aseptic loosening. Even though, recent efforts on using deep learning for detection of aseptic loosening based on the radiographs have shown great potential, they are mainly effective in detecting the late stages of loosening and early loosening detection remains challenging [12]. Today, no tools are available to effectively monitor the evolution of aseptic loosening in vivo.

In light of the preceding, we herein propose a novel diagnosis method by combining electrical impedance tomography (EIT) and self-sensing bone cement. More specifically, poly(methyl methacrylate) (PMMA) bone cement is made conductive by modification with low weight fractions of chopped carbon fibers (CF) above the percolation threshold (i.e. the lower limit of CF needed to form a well-connected network within the PMMA through which electrical current can propagate). Deformation and fracture alter the connectedness of the CF network and consequently manifest as a conductivity change. In other words, the material is piezoresistive, and conductivity changes can be used to monitor load transfer across the bone cement as a precursor to loss of implant fixation or outright failure of the bone cement.

Importantly, monitoring mechanical effects in piezoresistive materials via EIT has precedent in fields such as structural health monitoring (SHM) and nondestructive evaluation (NDE). For example, Tallman et al. modified glass fiber/epoxy laminates with carbon black (CB) fillers and used EIT to find impact damage [13]. The same group also used EIT to locate elastic deformations in a soft, carbon nanofiber (CNF)-modified polyurethane [14]. And, Gupta, Gonzalez, and Loh detected and localized artificially induced damage in multi-walled carbon nanotube (MWCNT)-modified cementitious composites via EIT [15]. In addition to these examples taken from SHM and NDE applications, a close relative of EIT, electrical capacitance tomography (ECT), has recently been studied for detecting and monitoring subcutaneous infection occurring at tissue-osseointegrated prosthesis interfaces via nanocomposite pH sensors [16]. These instances encourage the viability of the technique herein proposed.

II. CARBON FIBER-MODIFIED PMMA

*A. CF/PMMA Manufacturing*

Chopped CF (3.1 mm in length and 7 µm in diameter; Teijin Carbon, Tenax) were added to dental PMMA (Bosworth Fastray, Keystone Industries) to create a piezoresistive bone cement. Dispersion uniformity has a significant effect on both electrical and mechanical properties of the final piezoresistive bone cement. Therefore, to ensure good dispersion, CF were first dispersed into ethanol using a tip sonicator operating at 30 W and 20 kHz for one hour to break up clusters and individuate the CF. After sonication, the ethanol was removed by drying on a heated plate at 65 °C for two days to create a dry CF powder. The CF then was mechanically mixed with the dry polymer powder (20-to-120 µm diameter PMMA spheres coated with benzoyl peroxide, which acts as the polymerization initiator) using a mortar and pestle. Liquid methyl methacrylate (MMA) monomer with an accelerator N,N-dimethyl-p-toluidine (DmpT) was added (800 ml/g powder) to initiate a free-radical polymerization process and create the solid cement.

3D-printed molds were used to produce cylindrical samples for conductivity and stress-strain measurements and conical specimens which simulate the implant-cement-bone interaction. Cylindrical specimens measured 15 mm in length and 7.5 mm in diameter. Conical specimens (Fig. 1) were designed to apply compressive and shear stresses on a thin layer (~ 1 mm) of bone cement by sandwiching it between male and female components. The holes on the outer layer of the plastic surrogate expose bone cement to the water in the phantom tank later used for EIT. The conical section was modeled to simulate a physiological loading condition similar to standard hip implants [17].

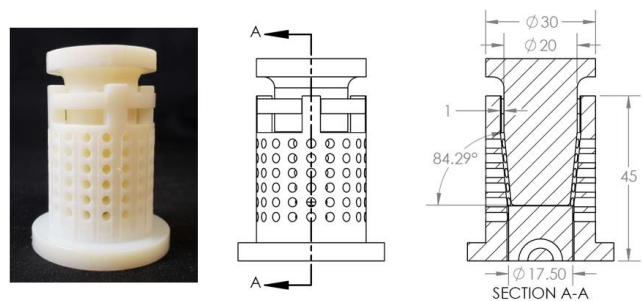

Figure 1. 3D-printed mold used to simulate the implant-cement-bone interaction and an overview and cross-sectioned mechanical drawing of the mold. Dimensions are in millimeters (mm). The bone cement is sandwiched between the male and female components.

*B. Electrical Properties of CF/PMMA*

Because conductivity changes are used as a mechanism for self-sensing in piezoresistive materials, it is important to thoroughly characterize the electrical properties of the CF/PMMA. For this, three properties were tested: direct current (DC) conductivity versus CF volume fraction, impedance versus frequency via electrical impedance spectroscopy (EIS), and DC conductivity change versus compressive load. For these tests, electrodes were attached by first polishing the ends of cylindrical specimens with 600-grit sandpaper, cleaning the specimen ends with ethanol, applying a layer of electrically conductive fast drying silver paint (Ted Pella 16040-30, Ted Pella Inc.), and applying copper tape electrodes atop the dried paste.

DC resistance was measured using the four-point probe method. From the measured resistance and specimen geometry, DC conductivity was calculated. In Fig. 2, a conductivity versus filler volume fraction behavior that is typical of percolation-dependent systems can be seen. That is, at the lowest concentrations of CF, the conductivity is low, followed by a significant jump as the filler concentration increases beyond the percolation threshold. Here, percolation is seen at approximately 1.5 vol.% CF.

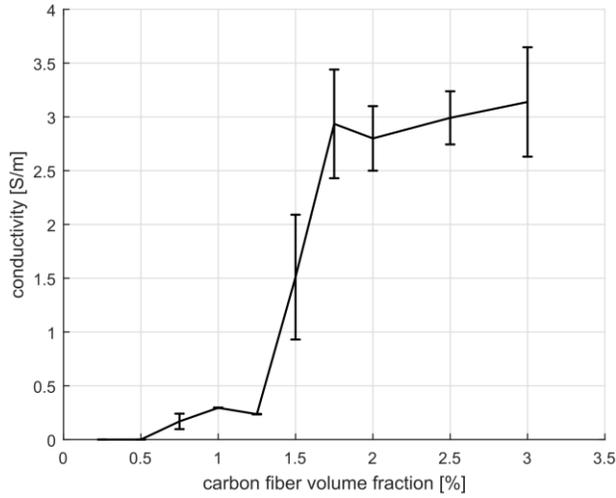

Figure 2. Conductivity of versus of CF volume fraction. Percolation behavior is seen around 1.5 vol.% CF. Specimens having 1.0, 1.5, and 2.0 vol.% are used later in EIT testing.

Next, EIS testing was conducted to assess the frequency-dependent electrical properties of the CF/PMMA. CF volume fractions 1.0, 1.5, and 2.0% were selected for these tests because they are also used in the EIT testing described later. Two specimens were tested for each weight fraction. The EIS data for these specimens is shown in Fig. 3. This data was generated by a Keysight E4990A impedance analyzer sweeping from 1 kHz to 10 MHz. The impedance analyzer was set to measure impedance magnitude and phase angle respectively denoted as $Z=Z'+jZ''$ and $\theta$ where $Z'$ is the real part of the impedance, $Z''$ is the complex part, and $j$ is the imaginary unit. From $Z$ and $\theta$, the real and complex parts can be calculated via Eqs. (1) and (2) below.

$$Z' = |Z|\cos(\theta) \quad (1)$$
$$Z'' = |Z|\sin(\theta) \quad (2)$$

A couple of interesting observations regarding the frequency-dependent behavior of the CF/PMMA can be made from Fig. 3. First, for the 1.0 vol.% CF specimen, a semi-circular shape can be seen. Similar behavior is widely reported for other percolation-dependent multi-phase material systems, notably nanocomposites with high aspect-ratio carbon nanofillers [18] [19]. Such behavior is routinely described by an equivalent circuit with a parallel resistor-capacitor component. Second, as the volume fraction is increased, the frequency response of the CF/PMMA changes drastically. Specifically, the semi-circular arcs seen at 1.0 vol.% disappear entirely and are replaced with nearly straight lines. Referring back to Fig. 2, this transition seemingly happens as the percolation threshold is crossed. The physical mechanisms of this transition in AC behavior, though interesting, are unknown at this point and a rigorous investigation is beyond the scope of this manuscript.

Last, conductivity was measured versus compressive load using a 10 kN capacity MTS 858 load machine. For this, cylindrical specimens were subjected to compressive loads while electrical resistance was measured simultaneously. Figure 4 shows the normalized conductivity versus strain for 1.0, 1.5, and 2.0 vol.% CF. The normalized conductivity is calculated by dividing the measured conductivity of each sample by the initial conductivity of the sample when no load is applied. From Fig. 4, the conductivity decreases with increasing compression. That is, the material exhibits negative piezoresistivity [20].

TABLE I
EQUIVALENT CIRCUIT PARAMETERS FOR EACH SPECIMEN

| | $R_p$ [Ω] | $R_s$ [Ω] | $C_p$ [×10⁻¹⁰ F] | $L_s$ [×10⁻⁶ H] |
|---|---|---|---|---|
| 1.0 vol.% specimen 1 | 432.50 | 83.26 | 18.66 | 1.66 |
| 1.0 vol.% specimen 2 | 601.52 | 92.74 | 16.46 | 2.10 |
| 1.5 vol.% specimen 1 | 183.89 | 53.31 | 32.54 | 1.42 |
| 1.5 vol.% specimen 2 | 240.89 | 64.83 | 22.10 | 1.49 |
| 2.0 vol.% specimen 1 | 36.08 | 72.63 | 2.37 | 1.19 |
| 2.0 vol.% specimen 2 | 38.15 | 138.56 | 2.35 | 1.13 |

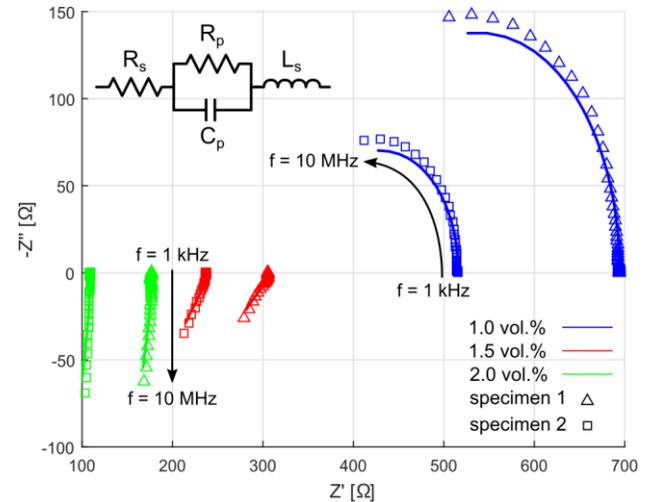

Figure 3. EIS data for 1.0, 1.5, and 2.0 vol.% CF. This behavior is well described by the equivalent circuit shown in the top-left of the plot. Experimental EIS data is shown as points and curve fits to the equivalent circuit are shown as solid lines.

### C. Mechanical Properties of CF/PMMA

To explore the effect of CF modification on PMMA mechanical properties, compressive stress-strain tests were conducted using the same MTS load frame. Figure 5 shows the stress-strain behavior for specimens with 0, 1.0, 1.5, and 2.0 vol.% CF. All of the samples show elastic-perfectly plastic behavior under compression. For the specimens tested, the elastic modulus and yield strength did not vary systematically with CF loading and were within 600-780 MPa and 27-35 MPa, respectively.

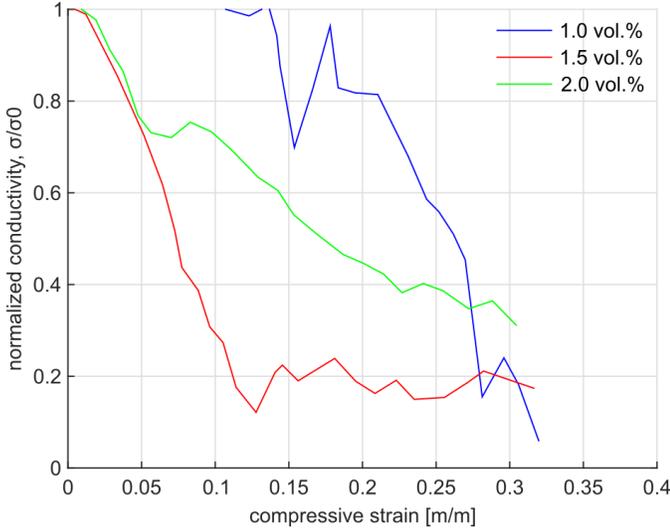

**Figure 1.** Normalized conductivity versus compressive strain. The conductivity of each specimen was divided by its initial conductivity for normalization.

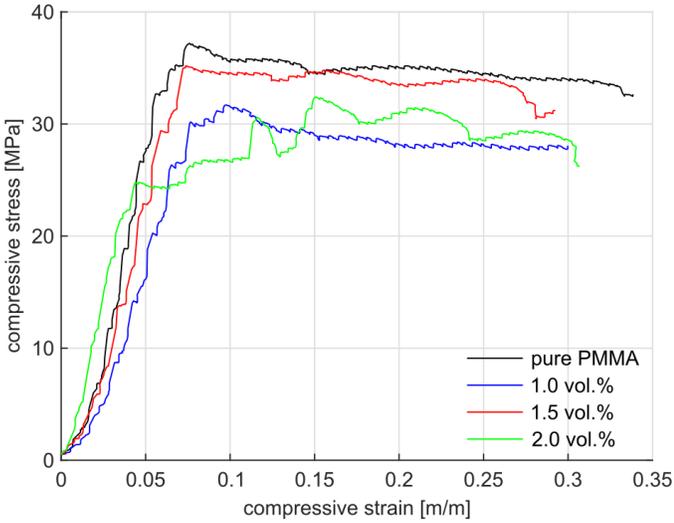

**Figure 2.** Stress-strain graph for pure PMMA and 1.0, 1.5, and 2.0 vol.% CF specimens.

## III. ELECTRICAL IMPEDANCE TOMOGRAPHY

### A. Forward Problem

Deformation-induced and damage-induced conductivity changes of the CF/PMMA bone cement interfacial layer were monitored via the EIT setup shown in Fig. 6. EIT estimates the conductivity distribution of a domain from a series of non-invasive current-voltage measurements collected at the domain boundary by minimizing the difference between a set of experimentally measured boundary voltages and another set of computationally predicted boundary voltages.

The process of computationally predicting voltages is referred to as the forward problem and begins with Laplace's equation for steady-state diffusion in the absence of internal current sources,

$$\nabla \cdot \sigma \nabla \phi = 0. \qquad (3)$$

Here, $\sigma$ is the conductivity distribution of the domain and $\phi$ is the domain potential. The complete electrode model boundary conditions are then enforced on Eq. (3) as shown in Eqs. (4) and (5). Equation (4) accounts for a voltage drop between the domain and the assumed to be perfectly conducting electrodes due to contact impedance. Equation (5) enforces conservation of charge. In these equations, $z_l$ is the contact impedance between the $l$th electrode and the domain, $\boldsymbol{n}$ is an outward pointing normal vector, $V_l$ is the voltage on the $l$th electrode, $E_l$ is the length of the $l$th electrode, and $L$ is the total number of electrodes.

$$\phi + z_l \sigma \nabla \phi \cdot \boldsymbol{n} = V_l \qquad (4)$$

$$\sum_{l=1}^{L} \int_{E_l} \sigma \nabla \phi \cdot \boldsymbol{n} \, dS_l = 0 \qquad (5)$$

Equations (3)-(5) are solved via the finite element method as shown in Eq. (6). In Eq. (6), $\boldsymbol{A}_M$ is the standard finite element stiffness matrix for steady-state diffusion, $\boldsymbol{\Phi}$ is a vector of domain potentials, $\boldsymbol{V}$ is a vector of electrode voltages, and $\boldsymbol{I}$ is a vector of currents injected at the electrodes. $\boldsymbol{A}_Z$, $\boldsymbol{A}_W$, and $\boldsymbol{A}_D$ are formed as respectively shown in Eqs. (7), (8), and (9) where $w_i$ is the $i$th finite element interpolation function. Two-dimensional triangles with linear interpolation are used in this work.

$$\begin{bmatrix} \boldsymbol{A}_M + \boldsymbol{A}_Z & \boldsymbol{A}_W \\ \boldsymbol{A}_W^T & \boldsymbol{A}_D \end{bmatrix} \begin{bmatrix} \boldsymbol{\Phi} \\ \boldsymbol{V} \end{bmatrix} = \begin{bmatrix} \boldsymbol{0} \\ \boldsymbol{I} \end{bmatrix} \qquad (6)$$

$$A_{Z\,ij} = \sum_{l=1}^{L} \int_{E_l} \frac{1}{z_l} w_i w_j \, dS_l \qquad (7)$$

$$A_{W\,li} = -\int_{E_l} \frac{1}{z_l} w_i \, dS_l \qquad (8)$$

$$\boldsymbol{A}_D = \text{diag}\left(\frac{E_l}{z_l}\right) \qquad (9)$$

### B. Inverse Problem

The EIT inverse problem is the process of recovering the conductivity distribution. Herein, we employ a one-step linearization process in which experimental voltages are collected at some initial time, again at some later time such as after some conductivity-changing event, and the conductivity change between these two times is sought. This method is used because it is more robust to noise and errors arising from factors such as discrepancies between electrode placement in the experiment and the model than absolute conductivity imaging methods. Mathematically, we seek a value of $\Delta\boldsymbol{\sigma}$ to minimize the difference between $\delta\boldsymbol{V} = \boldsymbol{V}(t_2) - \boldsymbol{V}(t_1)$ and $\boldsymbol{W} = \boldsymbol{F}(\boldsymbol{\sigma}_0 + \Delta\boldsymbol{\sigma}) - \boldsymbol{F}(\boldsymbol{\sigma}_0)$ in the least-squares sense as shown in Eq. (10).

$$\Delta\boldsymbol{\sigma}^* = \arg\min_{\Delta\boldsymbol{\sigma}} \frac{1}{2}(\|\boldsymbol{W}(\Delta\boldsymbol{\sigma}) - \delta\boldsymbol{V}\|_2^2) \qquad (10)$$

In the preceding equation, $\delta\boldsymbol{V}$ is the difference between experimentally collected voltages taken at times $t_1$ and $t_2$, $\boldsymbol{F}(\cdot)$ is a vector of electrode voltages predicted by the previously described forward problem for the conductivity distribution provided in its argument, $\boldsymbol{\sigma}_0$ is an estimate of the background conductivity, $\Delta\boldsymbol{\sigma}$ is the conductivity change between times $t_1$ and $t_2$, and $\Delta\boldsymbol{\sigma}^*$ is a conductivity change that satisfies the minimization. Note that $\boldsymbol{\sigma}$ has been boldfaced as a vector in anticipation of discretization by the finite element method. To solve this minimization, we approximate $\boldsymbol{F}(\boldsymbol{\sigma}_0 + \Delta\boldsymbol{\sigma})$ by using a Taylor series expansion and retain only the linear terms as shown in Eq. (11).

$$\boldsymbol{F}(\boldsymbol{\sigma}_0 + \Delta\boldsymbol{\sigma}) \cong \boldsymbol{F}(\boldsymbol{\sigma}_0) + \frac{\partial \boldsymbol{F}(\boldsymbol{\sigma}_0)}{\partial \boldsymbol{\sigma}} \Delta\boldsymbol{\sigma} \qquad (11)$$

By substituting Eq. (11) into $W$ and defining the sensitivity matrix as $J = \partial F(\sigma_0)/\partial \sigma$, we can recast Eq. (10) as shown in Eq. (12).

$$\Delta\sigma^* = \min_{\Delta\sigma \leq 0} \frac{1}{2}(\|J\Delta\sigma - \delta V\|_2^2 + \alpha\|L\Delta\sigma\|_2^2) \qquad (12)$$

Beyond eliminating $W$, two important differences between Eqs. (10) and (12) should be noted. First, we have added a constraint that the conductivity change be less than or equal to zero. This constraint is physically motivated by the fact that, as shown in Fig. 4, the CF/PMMA exhibited a conductivity loss with compression. And second, a regularization term, $L$, has been added. This regularization is necessary to recover a physically meaningful solution as the EIT inverse problem is underdetermined and ill-posed. Here, $L$ is the discrete Laplace operator and $\alpha$ is a scalar used to control the amount of regularization.

## IV. EXPERIMENTAL RESULTS AND DISCUSSION

### A. EIT Setup

A 16-electrode phantom tank measuring 133 mm in diameter and 80 mm deep was built for EIT testing and filled with deionized water. Deionized water was used to eliminate the effect of ion transport in the presence of direct current (DC) and to reduce oxidation. As shown in Fig. 6, the tank was built such that conical specimens acting as prosthetic surrogates could be loaded by an MTS load frame as EIT measurements are taken. A power source along with a linear regulator and an adjustable electrical resistor were used to inject 25 µA DC currents while voltages were measured via an Arduino Mega. Then the theoretical approach described in the previous section was used to calculate the EIT map.

Three conically shaped prosthetic surrogate specimens with 1.0, 1.5, or 2.0 vol.% CF cement layers were tested in this manner. One specimen per weight fraction was incrementally loaded up to 4000 N. EIT measurements were collected at each load step as the load was held constant. After loading up to 4000 N, each specimen was loaded until failure (occurring at approximately 5700, 6700, and 5700 N for the 1.0, 1.5, and 2.0 vol.% CF specimens, respectively). EIT measurements were again collected after each specimen failed.

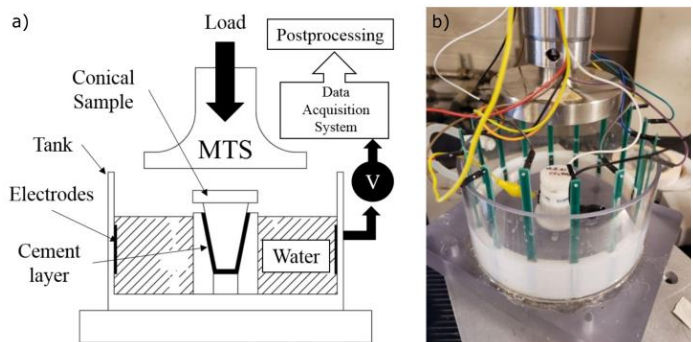

**Figure 3.** a) Schematic of the test setup. b) Image of experimental test phantom showing conical specimen in the center.

### B. Load Monitoring

EIT images for 1.0, 1.5, and 2.0 vol.% CF surrogates at increasing load levels can be seen in Fig. 7. In these images, EIT measurements at a load of 50 N were used as a baseline for difference imaging. This low level of loading was used as a baseline rather than a load-free condition because it is more representative of a person at rest (e.g. sitting or lying). EIT images corresponding to 450, 900, or 1350 N may be representative of quasi-static loads incurred by normal, low-stress activities such as walking or climbing stairs. Larger forces greater than 1350 N can be interpreted as being representative of dynamic loads incurred during more strenuous activities such as impact loads during running, jumping, or other forms of exercise. Consideration of these larger loads is important because, as mentioned in the introduction, the most at-risk populations for aseptic loosening are patients younger than 50 years old who lead more active lifestyles.

From Fig. 7, we can make a couple of important observations. First, we see that for each volume fraction considered, EIT can indeed adeptly identify the prosthetic location in the center of the phantom tank. Second, beyond location identification, the apparent magnitude of conductivity change depends on the load. That is, as the load magnitude increases, the conductivity change at the center of the phantom tank intensifies. This is important because it means that self-sensing bone cement and EIT can potentially be utilized for in vivo load monitoring.

Next, we calculate the average conductivity change as predicted by EIT in the center of the phantom, where the prosthetic surrogate is located, as a function of applied load. The diameter of the region over which the average conductivity change is calculated is 45 mm. This is shown in Fig. 8 where we see clear trend of increasing conductivity change magnitude with increasing load. A saturation effect can also be seen. That is, there is little change in $|\Delta\sigma|$ between 3100 and 4000 N. Saturation is a well-known compressive effect in piezoresistive polymer literature [21] and is a consequence of conductive fillers not being able to be pushed closer together indefinitely.

### C. Failure Detection

Next, we consider the potential of this method for failure detection. That is, each prosthetic surrogate is loaded past 4000 N until a gross failure event occurs. After failure, EIT measurements again are collected. Post-failure EIT images for each volume fraction are shown in Fig. 9. The 50 N case again is used as a baseline for difference imaging.

Several noteworthy observations can be made from Fig. 9. First, a conductivity loss due to failure can be clearly seen at the center of the phantom for both the 1.0 and 2.0 vol.% specimens. The conductivity loss is a consequence of the conductive bone cement fracturing and the ruptured volume being filled with less conductive deionized water. Second, despite being able to detect the occurrence of failure, EIT fails to capture the precise size or shape. However, the inability to capture precise shapes is a well-known limitation of EIT, especially at the center of the domain farthest away from electrodes [22]. And third, considering the top-middle plot of Fig. 9 which corresponds to the 1.5 vol.% CF specimen, we see that EIT seemingly fails to detect that the prosthetic has failed. That is, no conductivity change is detected by EIT at the center of the phantom where the prosthetic failure occurred. To better understand this, we need to recall that the EIT problem was formulated such that conductivity changes were constrained to be less than or equal to zero. If we remove this constraint and resolve the EIT inverse problem for the 1.5 vol.% specimen, the conductivity change shown in Fig. 10 is recovered.

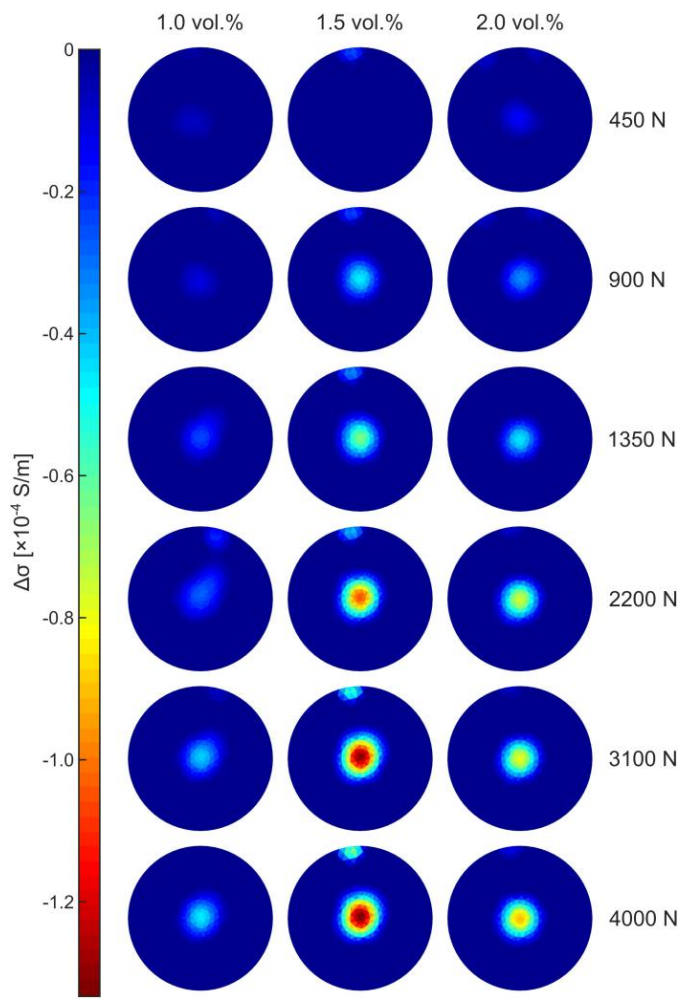

**Figure 4.** EIT-imaged conductivity changes of self-sensing bone cement as a function of applied load. All conductivity changes are constrained to be negative.

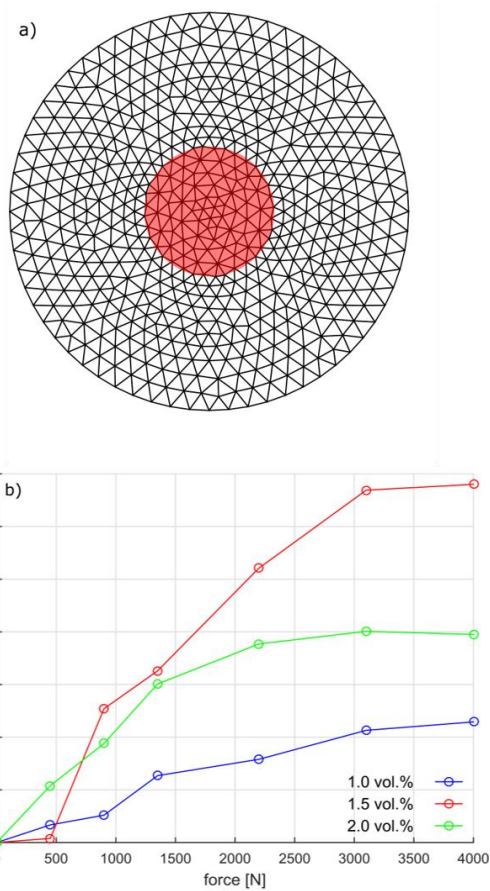

**Figure 5.** a) Finite element mesh used for EIT reconstructions. The average conductivity change of elements in the red region corresponding to the implant location is plotted as a function of the applied force. b) Average magnitude of conductivity change as imaged by EIT versus applied force.

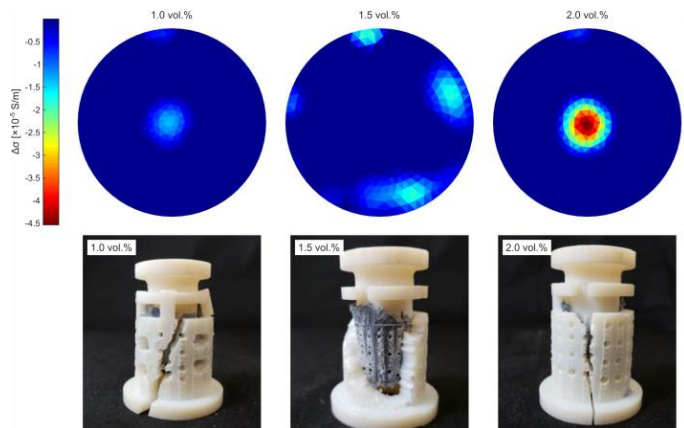

**Figure 6.** Top: EIT images after failure for each specimen. Failure-induced conductivity changes at the center of the phantom are clearly seen for the 1.0 and 2.0 vol.% CF specimens. Note these images again were formed with the constraint that the conductivity change be less than or equal to zero. Bottom: post-failure specimens.

Examining the EIT image in Fig. 10, we note a clear conductivity change at the center of the phantom. Further, this change is positive indicating that the region has become more conductive due to the failure event. To understand what is happening here, we need to examine the post-failure specimens as shown in Figs. 9 and 10. For specimens with 1.0 and 2.0 vol.% CF, both the bone cement and plastic surrogate cracked; however, for the specimen with 1.5 vol.% CF, only the plastic surrogate has broken. This significantly increased the surface area of CF/PMMA exposed to the water. Additionally, the bone cement did not actually break. Because the CF/PMMA is more conductive than the deionized water and a greater amount of it is in direct contact with the water after the failure event, there is an easier path for electric current to flow and hence an overall higher apparent conductivity in the center of the phantom tank just as predicted by EIT. This failure can be thought of as being similar to de-bonding between the PMMA and bone or between the PMMA and implant. Even though this particular failure mode of the 1.5 vol.% specimen was not by design, the results are nonetheless noteworthy because it indicates that distinguishing between failure types may be possible. That is, a fracture or breakage of the cement resulted in a conductivity loss whereas a de-bonding event increased the EIT-imaged conductivity change.

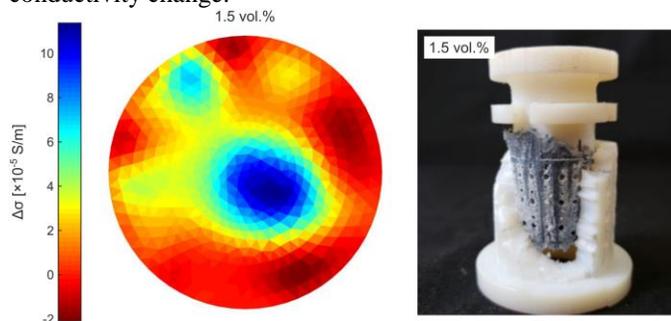

**Figure 7.** EIT image of the 1.5 vol.% specimen without any constraints on the conductivity change. EIT detects a conductivity increase due to a de-bonding failure event (as opposed to CF/PMMA fracture for the 1.0 and 2.0 vol.% specimens).

*D. Discussion*

The preceding results suggest that interfacial monitoring via piezoresistive bone cement and EIT can potentially be used for detection of aseptic loosening. That is, EIT was able to successfully identify and distinguish between increasing load levels and detect gross implant failure. This success is noteworthy because EIT has several important advantages in this application compared to more common modalities such as radiographs. First, unlike ionizing radiation, low levels of electrical current are physiologically benign. Second, the equipment required for EIT is cost-effective, generally only requiring precision current supplies, voltmeters, and modest computational power. And third, EIT has high temporal resolution; optimized systems can generate images in nearly real time. Combined, these advantages mean that EIT could potentially be used continuously (e.g. via a wearable cuff with built-in electrodes) on high-risk patients to provide real-time information on in vivo implant fixation, load transfer, and failure. Not only is this a potential clinical boon, but it could also be of significant research relevance by providing real-time insight on factors which precipitate loss of implant fixation.

Despite the successes of this preliminary study, several important limitations and directions for future refinement should be discussed. First, the 450 N loading case is only faintly visible for 1.0 and 2.0 vol.% CF and invisible for 1.5 vol.% CF. Tuning and optimization of the bone cement's piezoresistive properties and the EIT formulation could improve sensitivity to low loads. Second, the failure events detected in this preliminary study were quite exaggerated. A sensitivity study should therefore be conducted in the future in order to understand the lower limits of failure (i.e. cement cracking and cement-to-implant or cement-to-bone de-bonding) detection. Third, this work was conducted in a phantom tank. Real physiology is obviously much more complex with heterogeneous and anisotropic conductivity due to native tissues. Future work should consequently utilize cadaver testing. And fourth, CF was used for proof-of-concept demonstration herein. Physiological compatibility is an important issue that should be considered in future work on self-sensing bone cement. For example, much previous work has explored the effect of nanoscale additives such as carbon nanotubes (CNT) and silver nanowires (AgNW) on the electrical and mechanical performance and physiological compatibility of polymer nanocomposites [23] [24] [25] [26] [27]. Moreover, nanocomposites using CNTs have shown biocompatibility [28] [29].

## V. CONCLUSIONS

In conclusion, we have presented a novel approach to monitoring interfacial load transfer and failure in cemented joint replacements. This approach is based on monitoring the conductivity changes of conductive filler-modified PMMA with EIT and is capable of detecting both physiological-level loading and gross implant failure. Further, we distinguished between increasing load magnitudes and failure types. Based on these results, several suggestions for future work are made including enhancing sensitivity to low loads, identifying the lower limit of failure detectability, exploring feasibility in the presence of native tissue, and transitioning to physiologically compatible nanofillers.